# Tuning the electronic structure of α-antimonene monolayer through interface engineering

Zhi-Qiang Shi,[†] Huiping Li,[‡,§] Cheng-Long Xue,[†] Qian-Qian Yuan,[†] Yang-Yang Lv,[⊥] Yong-Jie Xu,[†] Zhen-Yu Jia,[†] Libo Gao,[†,∥] Yanbin Chen,[†,∥] Wenguang Zhu,[‡,§,*] Shao-Chun Li[†,∥,#,*]

[†] *National Laboratory of Solid State Microstructures, School of Physics, Nanjing University, Nanjing 210093, China*

[‡] *International Center for Quantum Design of Functional Materials (ICQD), Hefei National Laboratory for Physical Sciences at the Microscale, and Synergetic Innovation Center of Quantum Information and Quantum Physics, University of Science and Technology of China, Hefei, Anhui 230026, China*

[§] *Key Laboratory of Strongly-Coupled Quantum Matter Physics Chinese Academy of Sciences, School of Physical Sciences, University of Science and Technology of China, Hefei, Anhui 230026, China*

[⊥] *National Laboratory of Solid State Microstructures, Department of Materials Science and Engineering, Nanjing University, Nanjing 210093, China*

[∥] *Collaborative Innovation Center of Advanced Microstructures, Nanjing University, Nanjing 210093, China*

[#] *Jiangsu Provincial Key Laboratory for Nanotechnology, Nanjing University, Nanjing 210093, China*

These authors contributed equally: Zhi-Qiang Shi, Huiping Li, Cheng-Long Xue, Qian-Qian Yuan. Correspondence and requests for materials should be addressed to Wenguang Zhu (email: wgzhu@ustc.edu.cn); Shao-Chun Li (email: scli@nju.edu.cn).

**ABSTRACT:** The interfacial charge transfer from the substrate may influence the electronic structure of the epitaxial van der Waals (vdW) monolayers and thus their further technological applications. For instance, the freestanding Sb monolayer in puckered honeycomb phase (α-antimonene), the structural analog of black phosphorene, was predicted to be a semiconductor, but the epitaxial one behaves as a gapless semimetal when grown on the $T_d$-WTe$_2$ substrate. Here, we demonstrate that interface engineering can be applied to tune the interfacial charge transfer and thus the electron band of epitaxial monolayer. As a result, the nearly freestanding (semiconducting) α-antimonene monolayer with a band gap of ~170 meV was successfully obtained on the SnSe substrate. Furthermore, a semiconductor-semimetal crossover is observed in the bilayer α-antimonene. This study paves the way towards modifying the electron structure in two-dimensional vdW materials through interface engineering.

The fundamental physics at the interface plays a significant role in various fields, such as electronics [1-4], optoelectronic [5,6], catalysis [7-9], solar energy conversion [10,11], etc. In contrast to a strongly coupled interface that usually modulates the structure of epitaxial layer [12-23], charge transfer is the dominant phenomenon at a weakly coupled interface that effectively dopes electrons or holes to the epitaxial layers [24-26]. For van der Waals (vdW) metal-semiconductor junctions, the interlayer charge transfer occurs upon contact in order to align the Fermi level [27-31]. Moreover, exotic physics can be also induced by the interfacial charge transfer, as particularly exemplified by the discovery of enhanced superconducting transition in the epitaxial FeSe monolayer on $SrTiO_3$ substrate [32-38]. However, for the case of narrow gap semiconductors, it is fatal that the intrinsic band gap is suppressed due to the interfacial charge transfer, such as the epitaxial stanene on $Bi_2Te_3$ [25] and α-antimonene on $T_d$-$WTe_2$ [39]. Therefore, to finely tune the electronic structure of the epitaxial monolayer is challenging due to the difficulty of simultaneously considering both of lattice mismatch and electron hybridization.

Two-dimensional (2D) monolayers of group IV elements have been predicted to be the candidates hosting the quantum spin Hall effect [40-42]. However, a sizeable gap opened by spin-orbit coupling can be easily damaged due to the substrate effect [25,43]. The black phosphorene analog of antimony, namely α-antimonene, was theoretically predicted to be a narrow gap semiconductor [44,45] with high carrier mobility comparable to graphene [46]. Besides, it was also predicted to be an elemental ferroelectric material [47] and thermoelectric material [46]. Even though the large-scale and high-quality α-antimonene monolayer has been recently realized on the $T_d$-$WTe_2$ substrate [39], it exhibits a metallic nature, instead of a semiconductor as predicted theoretically [45], which can be possibly ascribed to the coupling to the substrate [39]. To date, the α-antimonene monolayer with semiconducting band structure has not been reported.

In this work, we explored ways to tune the electronic structure of α-antimonene monolayer through interface engineering. The α-antimonene monolayer was grown on different substrates by using molecular beam epitaxy (MBE), and characterized by scanning tunneling microscopy/spectroscopy (STM/STS) in combination with density functional theory (DFT) calculations. Through optimizing the growth conditions, antimonene adopts a two dimensional vdW growth mode on both SnSe and MoTe$_2$ substrates. As a result, the semiconducting α-antimonene monolayer was realized on the SnSe substrate, which exhibits a band gap of ~170 meV, in good agreement with the DFT calculations for the freestanding one. A semiconductor-semimetal transition was also discovered as the thickness is increased to bilayer. When grown on the T$_d$-MoTe$_2$ substrate, the spectroscopic measurement indicated that the α-antimonene monolayer was metallic. We found that the electronic structure of the epitaxial α-antimonene monolayer is largely determined by the magnitude of the interfacial charge transfer. Unexpectedly, the lattice mismatch induced strain effect plays a non-dominant role in leading to the formation of a band gap.

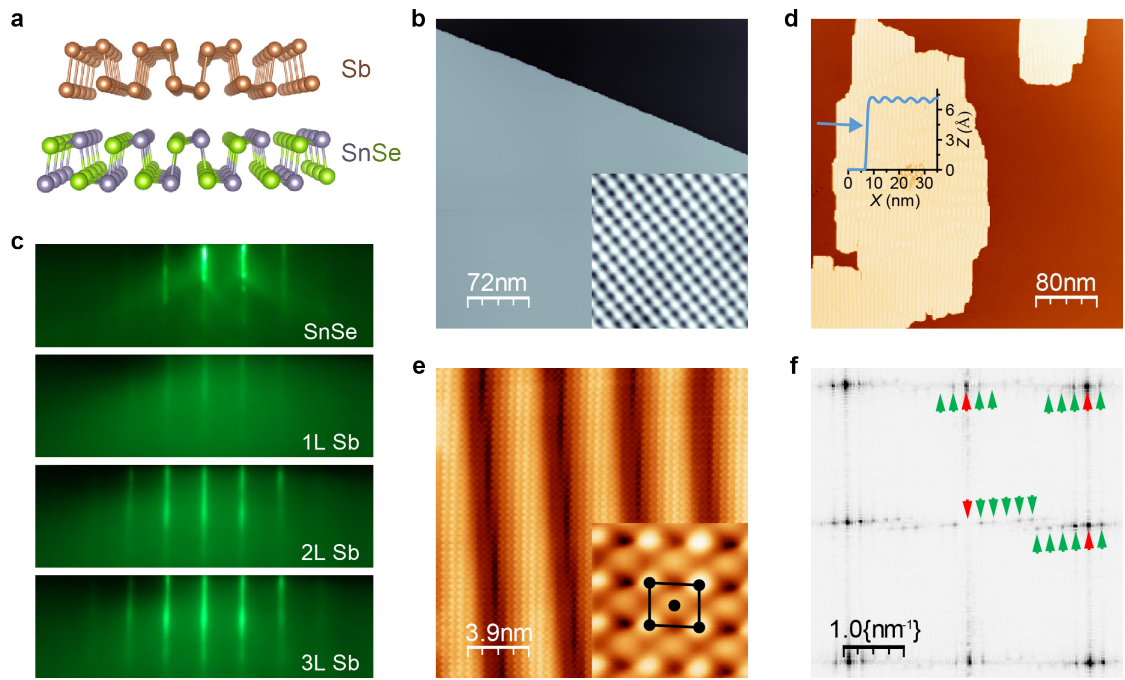

**Figure 1.** The epitaxial growth and STM characterization of ~0.5 ML antimony deposited on SnSe substrate. (a) The schematics of epitaxial growth of α-antimonene on orthorhombic SnSe substrate. (b) The topographic image (360 × 360 nm$^2$) of the bare SnSe substrate. $U$ = +4 V, $I_t$ = 100 pA. The inset shows the atomically resolved STM image (5 × 5 nm$^2$). $U$ = +800 mV, $I_t$ = 50 pA. (c) The RHEED patterns recorded before and after the deposition of 1-3 ML Sb on the SnSe substrate. (d) The STM morphology image (400 × 400 nm$^2$) of ~0.5 ML Sb deposited on SnSe at 400 K. The inset shows the step height of the α-antimonene monolayer (~6.8 Å). $U$ = +3 V, $I_t$ = 100 pA. (e) The atomically resolved STM image (19.5 × 19.5 nm$^2$) taken on the α-antimonene monolayer. $U$ = -880 mV, $I_t$ = 5 nA. The inset shows the enlarged image (1.5 × 1.5 nm$^2$) of (e). The 1 × 1 unit cell is marked by the black rectangle. $U$ = -1 V, $I_t$ = 1 nA. (f) The fast Fourier transform (FFT) of (e), the red and green arrows mark the Bravais lattice of 1 × 1 α-antimonene and the reciprocal vectors of moiré pattern, respectively.

Figure 1a schematically illustrates the epitaxial growth of the α-antimonene monolayer on the SnSe substrate. Bulk SnSe is a semiconductor with a band gap of ~0.8-0.9 eV [48, 49]. It adopts the orthorhombic crystalline structure (space group: Pnma), which resembles to the α-antimonene (space group: Pmn2$_1$) if different type of atoms is disregarded [48, 49]. As shown in the STM topographic image (Fig. 1b), the bare surface of the cleaved SnSe substrate is composed of the atomically flat terraces in micrometer scale. The atomic resolution image, the inset to Fig. 1b, shows clearly the topmost Sn atoms, in good agreement with previous reports [49]. During the epitaxial growth of antimony on SnSe, the surface morphology was *in situ* monitored via RHEED patterns. Figure 1c shows the RHEED patterns recorded on the bare SnSe and the epitaxial antimonene films with nominal coverages from 1 to 3 monolayers (MLs), respectively. The streaks for the bare SnSe again indicate the high crystalline quality and the atomically flat surface, and those for the antimony films indicate that Sb takes a layer-by-layer growth mode on SnSe.

Figure 1d shows the surface morphology of ~0.5 ML Sb deposited on SnSe substrate. The larger scale SEM images are shown in Fig. S1, Supplementary Information. The surface is characterized as the homogeneous antimonene islands in micrometer scale and the exposed SnSe substrate. Line-scan profile measured across the step edge, as plotted in the inset to Fig. 1d, indicates the thickness of these islands is ~6.8 Å, comparable to the α-antimonene monolayer [44]. High-resolution STM image, Fig. 1e, shows the high quality antimonene monolayer without obvious defects. The notable modulation, as identified in Fig. 1e, is attributed to the moiré pattern formed through lattice superimposition between the epitaxial antimonene and SnSe substrate. The inset image to Fig. 1e shows clearly the surface atoms that match very well to the atomic termination of α-antimonene. The fast Fourier transform (FFT) of the antimonene topography, Fig. 1f, shows the Bravais lattice of 1 × 1 antimonene and the reciprocal vectors of the moiré pattern, as marked by red and green arrows respectively. The extracted in-plane lattice constants are ~4.29 Å and ~4.76 Å along the zigzag and armchair direction respectively, close to freestanding α-antimonene ($a$ = 4.36 Å, $b$ = 4.74 Å) [46]. The estimated lattice strain is as small as ~ -1.6% (compressed) along the zigzag, and ~ +0.4% (stretched) along the armchair direction. Different from the previous study of α-antimonene on $T_d$-WTe2 substrate where a $\sqrt{2} \times \sqrt{2}$ superstructure is formed [39], the α-antimonene monolayer grown on SnSe substrate exhibits only the 1 × 1 periodicity. Raman measurements (see Fig. S2, Supplementary Information) suggest that the epitaxial antimonene film is kept in the α-phase (puckered honeycomb structure) at least for up to ~15 nm thick.

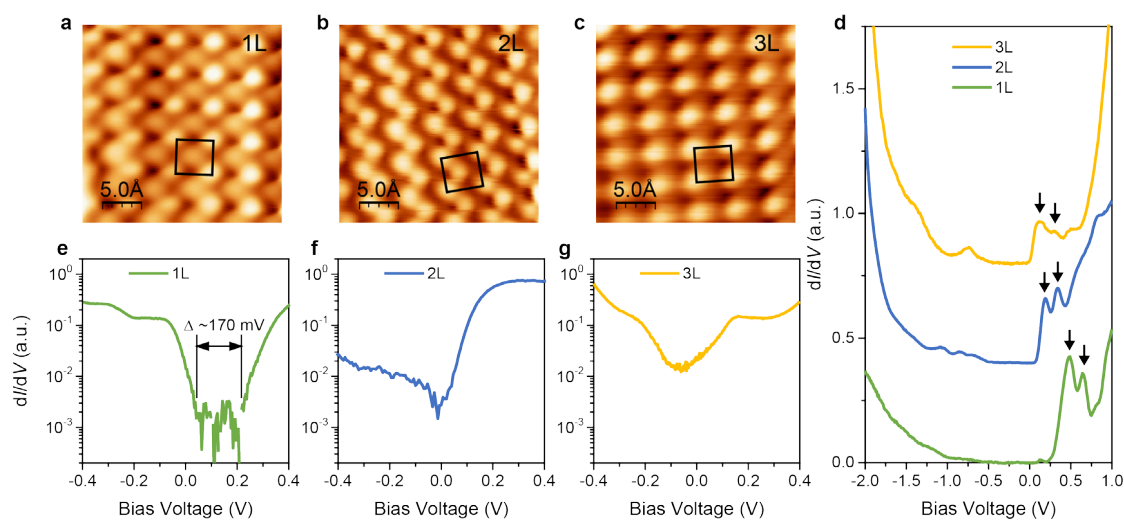

**Figure 2.** The STM/STS characterization of 1-3 ML α-antimonene. (a-c) The atomically resolved STM images (2.5 × 2.5 nm$^2$) taken on the monolayer, bilayer and tri-layer α-antimonene, respectively. The black rectangles mark the 1 × 1 unit cell. $U$ = -500 mV, $I_t$ = 4 nA. (d) Thickness-dependent dI/dV spectra (from -2.0 V to +1.0 V) taken on the antimonene films from 1-3 ML. The characteristic peaks are marked by black arrows in each spectrum. The spectra are offset vertically for clarify. (e-g) The logarithmic format of the dI/dV spectra near the Fermi energy

(from -0.4 V to +0.4 V) for the 1-3 ML antimonene films. The bandgap of α-antimonene monolayer is ~170 meV.

To investigate the electronic structure of the epitaxial α-antimonene, dI/dV spectroscopy (STS) was measured. Prior to STS measurement, the atomically resolved images, as shown in Figs 2a-c, were collected to verify the puckered honeycomb structure for the monolayer, bilayer and tri-layer antimonene. Figure 2d shows the large-scale dI/dV spectra (from -2 V to +1 V) taken on the α-antimonene from 1 to 3 MLs. The characteristic features, as marked by the black arrows in Fig. 2d, evolve gradually with thickness due to the interlayer coupling. The logarithmic format of the dI/dV spectra near the Fermi energy (from -0.4 V to +0.4 V) are plotted accordingly in Figs 2e-g. Figure 2e shows that the antimonene monolayer has a bandgap of ~170 meV, comparable with the previous prediction of ~190 meV for the freestanding α-antimonene monolayer [45] and our calculated result in Fig. S3, Supplementary Information. Both of the dI/dV spectra taken on the bilayer and tri-layer antimonene, as plotted in Figs 2f, g, show a metallic density of state, demonstrating that a semiconductor-semimetal crossover occurs at 2 MLs. These spectroscopic results are consistent with the theoretical prediction for the freestanding antimonene [44, 45] and our calculated results of epitaxial α-antimonene in Fig. S3, Supporting Information. According to both of the experimentally determined lattice constants and electronic structure, it is therefore concluded that the epitaxial α-antimonene monolayer on SnSe substrate can be considered as nearly freestanding.

To further explore the interfacial impact to the electronic structure of the α-antimonene, we adopted another substrate of $T_d$-MoTe$_2$ to grow antimony monolayers. Single crystal $T_d$-MoTe$_2$ is a semimetal, and its crystal structure (space group: Pmn2$_1$, $a$ = ~6.33 Å and $b$ = ~3.47 Å) is largely different from that of SnSe [50]. Figure 3a shows the surface of ~2.5 ML antimony deposited on $T_d$-MoTe$_2$ substrate at 78 K followed by annealing to room temperature. The regions of monolayer, bilayer and tri-layer antimonene are identified, as marked by green, blue and orange arrows, respectively. All of these regions crystallize in the α-phase, as verified by the atomically resolved images (Figs 3b-d) and the large-scale STS spectra (Fig. 3e). Moreover, the characteristic features in the dI/dV spectra, as marked by the black arrows in Fig. 3e, are similar to those for the α-antimonene on SnSe (Fig. 2d), except for a systematic shift of ~ +750 mV towards higher bias voltage, which may be due to the different charge transfer occurring at the interface.

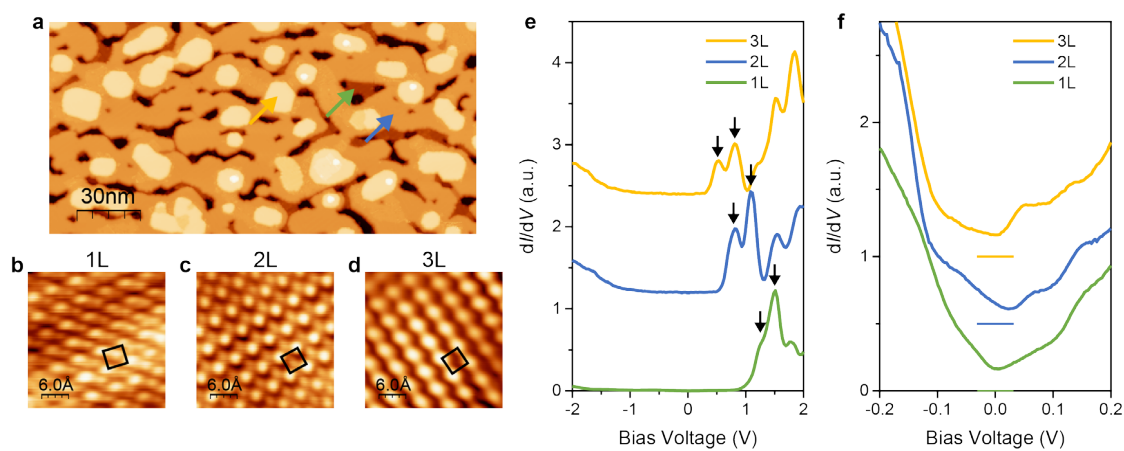

**Figure 3.** The low temperature growth and STM characterization of ~2.5 ML Sb deposited on Td-MoTe$_2$ substrate. (a) STM image (200 × 100 nm$^2$) of ~2.5 ML Sb deposited on MoTe$_2$ substrate at 78 K, followed by thermally annealing to room temperature. The green, blue and orange arrows mark the regions of 1-3 ML antimonene films, respectively. $U$ = +1 V, $I_t$ = 100 pA. (b-d) The atomically resolved STM images (3 × 3 nm$^2$) of 1-3 ML α-antimonene, and the 1 × 1 unit cell are marked by black rectangle accordingly. $U$ = +10 mV, $I_t$ = 100 pA. (e) Thickness-dependent dI/dV spectra (from -2.0 V to +2.0 V) taken on the 1-3 ML antimonene/MoTe$_2$ films, the characteristic peaks are marked by black arrows in each spectrum. The spectra are offset vertically for clarify. (f) The enlarged STS of (e) from -0.2 V to +0.2 V. The green, blue and orange base lines mark the zero point of the dI/dV intensity accordingly. The spectra are offset vertically for clarify.

The dI/dV spectra taken near the Fermi energy, Fig. 3f (from -0.2 V to +0.2 V), indicate the α-antimonene/MoTe$_2$ monolayer is instead a metal, and both of the bilayer and tri-layer antimonene/MoTe$_2$ are metallic as well. The lattice constants of the α-antimonene/MoTe2 monolayer are determined to be ~4.46 Å along the zigzag and ~ 4.87 Å along the armchair direction. The tensile strain is estimated as ~+2.3% (~+2.8%) along the zigzag (armchair) direction, which is much larger than that in the antimonene on SnSe.

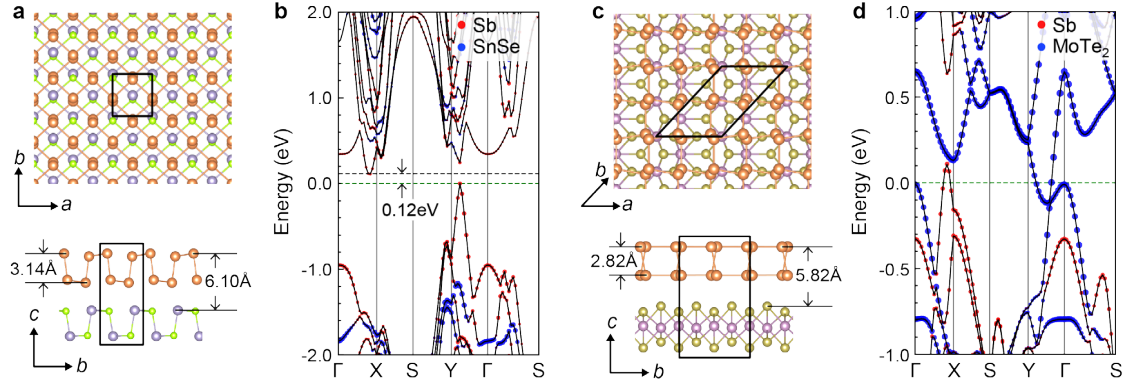

**Figure 4.** Structural configuration and corresponding projected electronic band structure for α-antimonene on SnSe and MoTe2 substrate. (a) Top and side view of DFT-optimized atomic structure model of α-antimonene on SnSe substrate, the commensurate unit cell is indicated by the black box. (b) HSE06 band structure of α-Sb/SnSe system, with an energy gap of 0.12 eV. (c) Top and side view of DFT-optimized atomic structure model of α-antimonene on MoTe2 substrate, with the $\sqrt{2}\times2$-Sb/$1\times\sqrt{5}$-MoTe2 supercell used for calculation indicated. (d) HSE06 band structure of α-Sb/MoTe2 system. The Fermi level shown as the green dashed lines in (b) and (d) are set to zero in energy.

To help understand the experimental results, we performed density functional theory (DFT) calculations for α-antimonene on both SnSe and MoTe2 substrates, as shown in Fig. 4. The α-antimonene on SnSe has a commensurate lattice, with the lattice constants of $a$ = 4.25 Å and $b$ = 4.61 Å in the fully optimized structure. In comparison to the lattice of free-standing α-antimonene monolayer (see Fig. S3, Supplementary Information), the substrate applies a compressive strain of 2.5% to the zigzag direction and 2.7% to the armchair direction of the adlayer, and the adlayer remains an asymmetric washboard structure, as shown in Fig. 4a. The changes to the band structure for α-antimonene under such small strain are minimal [46]. Due to the large band gap of SnSe and proper band alignment, the energy gap of α-antimonene is preserved on the SnSe substrate, as shown in Fig. 4b. The minimal lattice mismatch results in the formation of nearly freestanding α-antimonene monolayer with a semiconducting band structure, in good agreement with the experimental results. For the α-antimonene on $T_d$-MoTe$_2$, the calculated lattice parameters are stretched by 3.2% along the zigzag direction and 3.9% along the armchair direction in comparison to the free-standing α-antimonene monolayer, and the adlayer changed towards almost a symmetric washboard-like structure, as shown in Fig. 4c. According to previous DFT study [46], such tensile strain leads to the increase of the band gap, as illustrated by the extracted Sb part of the projected band structure in Fig. 4d. However, with the $T_d$-MoTe$_2$ substrate, the Fermi level is shifted downwards into valence bands of the α-antimonene, accompanied by electron transfer from the epitaxial monolayer to the substrate, resulting in a hole-doped character and the experimentally observed metallic STS signal. Despite of the inevitable deviation of the structural parameters and energy gaps between theoretical and experimental value, the calculated band structures qualitatively reproduce the experimental observations.

We have successfully realized the narrow gap semiconducting α-antimonene monolayer through interface engineering. It is also demonstrated to be a practically feasible way to tune the band gap of the vdW epitaxial monolayer through interface engineering. Superior to black phosphorene, α-antimonene monolayer remains stable in air. Regarding the high carrier mobility as theoretically predicted, the semiconducting α-antimonene monolayer shows great potential as a new candidate for future nanoelectronic devices.

## METHODS

### Characterizations

STM and MBE: The antimonene films were grown in ultrahigh vacuum (UHV) chamber with a base pressure of $1 \times 10^{-10}$ Torr. The substrate of SnSe and Td-MoTe$_2$ single crystals were *in situ* cleaved in UHV. Prior to the sample growth, the surface morphology of the substrates was checked by STM and the reflection high-energy electron diffraction (RHEED). The high quality antimony (99.999%) was thermally evaporated from a standard Knudsen cell, and the flux was kept at ~ 0.3 monolayer (ML) /min. All the STM measurements were performed in the MBE-STM combined system (Unisoku USM-1500) at 78 K with a PtIr tip. Constant current mode was utilized to collect the topographic images. STS were taken with a lock-in amplifier and the ac modulation of 10-20 mV at 978 Hz was applied.

### Computational methods

DFT calculations: The first-principles DFT calculations were performed by using the Vienna *ab initio* simulation package (VASP) [51], which is based on the projector-augmented wave (PAW) method [52, 53]. The exchange-correlation functional was treated using the Perdew-Burke-Ernzerhof (PBE) [54] parameterization of generalized gradient approximation (GGA) for structural relaxations and total energy calculations. The band structures were calculated by using the hybrid functional of Hyed-Scuseria-Ernzerhof (HSE06) [55]. A cutoff energy of 350 eV for the plane wave basis set was used. The Brillouin zone is sampled by a $15 \times 12 \times 1$ and $12 \times 9 \times 1$ Γ-centered *k*-mesh for the $1 \times 1$-Sb/$1 \times 1$-SnSe and $\sqrt{2} \times 2$-Sb/$1 \times \sqrt{5}$-MoTe$_2$ system, respectively. And a vacuum layer over 15 Å was added to all slabs to avoid the coupling between periodic images. Optimized structures were achieved when forces on all the atoms were smaller than 0.01 eV/Å. The van der Waals corrections were treated by the Grimme's DFT-D2 method [56] in the heterostructure calculations.

## ASSOCIATED CONTENT

**Supporting Information**.

This material is available free of charge via the internet at http://pubs.acs.org. Scanning electron microscopy data of Sb/SnSe monolayers (Figure S1), Raman spectra of Sb/SnSe and Sb/MoTe$_2$ monolayers (Figure S2), DFT calculated atomic and band structure of α-antimonene monolayer, bilayer and trilayer (Figure S3).

## ACKNOWLEDGEMENTS

This work was financially supported by the National Natural Science Foundation of China (Grants Nos. 11774149, 11790311, 11674299, 11634011, 51872134, 11574131, 51902152, 11674154 and 11761131010), the Foundation for Innovative Research group of the National Natural Science Foundation of China (No. 51721001); the National Key Research and Development Program of China (Grant Nos. 2017YFA0204904 and 2019YFA0210004), the Strategic Priority Research Program of Chinese Academy of Sciences (Grant No. XDB30000000), Anhui Initiative in Quantum Information Technologies (Grant No. AHY170000), and the Fundamental Research Funds for the Central Universities (Grant Nos. WK2340000082 and WK2060190084).

## AUTHORS INFORMATION

**Corresponding authors**

**Wenguang Zhu** – *International Center for Quantum Design of Functional Materials (ICQD), Hefei National Laboratory for Physical Sciences at the Microscale, and Synergetic Innovation Center of Quantum Information and Quantum Physics, Key Laboratory of Strongly-Coupled Quantum Matter Physics Chinese Academy of Sciences, School of Physical Sciences, University of Science and Technology of China, Hefei, Anhui 230026, China;* Email: wgzhu@ustc.edu.cn

**Shao-Chun Li** – *National Laboratory of Solid State Microstructures, School of Physics, Collaborative Innovation Center of Advanced Microstructures, Jiangsu Provincial Key Laboratory for Nanotechnology, Nanjing University, Nanjing 210093, China;* orcid.org/0000-0001-9818-4255; Email: scli@nju.edu.cn

**Authors**

**Zhi-Qiang Shi** – *National Laboratory of Solid State Microstructures, School of Physics, Nanjing University, Nanjing 210093, China*

**Huiping Li** – *International Center for Quantum Design of Functional Materials (ICQD), Hefei National Laboratory for Physical Sciences at the Microscale, and Synergetic Innovation Center of Quantum Information and Quantum Physics, Key Laboratory of Strongly-Coupled Quantum Matter Physics Chinese Academy of Sciences, School of Physical Sciences, University of Science and Technology of China, Hefei, Anhui 230026, China*

**Cheng-Long Xue** – *National Laboratory of Solid State Microstructures, School of Physics, Nanjing University, Nanjing 210093, China*

**Qian-Qian Yuan** – *National Laboratory of Solid State Microstructures, School of Physics, Nanjing University, Nanjing 210093, China*

**Yang-Yang Lv** – *National Laboratory of Solid State Microstructures, Department of Materials Science and Engineering, Nanjing University, Nanjing 210093, China*

**Yong-Jie Xu** – *National Laboratory of Solid State Microstructures, School of Physics, Nanjing University, Nanjing 210093, China*

**Zhen-Yu Jia** – *National Laboratory of Solid State Microstructures, School of Physics, Nanjing University, Nanjing 210093, China*

**Libo Gao** – *National Laboratory of Solid State Microstructures, School of Physics, Collaborative Innovation Center of Advanced Microstructures, Nanjing University, Nanjing 210093, China*

**Yanbin Chen** – *National Laboratory of Solid State Microstructures, School of Physics, Collaborative Innovation Center of Advanced Microstructures, Nanjing University, Nanjing 210093, China*

**Author contributions**

Z.-Q.S., H.L., C.-L.X. and Q.-Q.Y. contributed equally to this work. S.-C.L. conceived the project. Z.-Q.S., C.-L.X. and Q.-Q.Y. grew the antimonene films and carried out STM experiments with the assistance of Y.-J.X. and Z.-Y.J. H.L. and W.Z. carried out the theoretical calculations. Y.-Y.L. and Y.C. provided the single-crystal substrates. L.G. performed the Raman measurements. Z.-Q.S. and S.-C.L. wrote the manuscript with the input from H.L. and W.Z. All authors discussed the results and commented on the manuscript.

**Notes**

The authors declare no competing financial interest.

**REFERENCES**

(1) McKee, R. A.; Walker, F. J.; Buongiorno Nardelli, M.; Shelton, W. A.; Stocks, G. M. The interface phase and the Schottky barrier for a crystalline dielectric on silicon. *Science* **2003**, 300, 1726.

(2) Ohtomo, A.; Hwang, H. Y. A high-mobility electron gas at the LaAlO$_3$/SrTiO$_3$ heterointerface. *Nature* **2004**, 427, 423.

(3) Reyren, N.; Thiel, S.; Caviglia, A. D.; Fitting Kourkoutis, L.; Hammerl, G.; Richter, C.; Schneider, C. W.; Kopp, T.; Rüetschi, A.-S.; Jaccard, D.; Gabay, M.; Muller, D. A.; Triscone, J.-M.; Mannhart, J. Superconducting interfaces between insulating oxides. *Science* **2007**, 317, 1196.

(4) Soumyanarayanan, A.; Reyren, N.; Fert, A.; Panagopoulos, C. Emergent phenomena induced by spin-orbit coupling at surfaces and interfaces. *Nature* **2016**, 539, 509.

(5) Hong, X.; Kim, J.; Shi, S.-F.; Zhang, Y.; Jin, C.; Sun, Y.; Tongay, S.; Wu, J.; Zhang, Y.; Wang, F. Ultrafast charge transfer in atomically thin MoS$_2$/WS$_2$ heterostructures. *Nat Nanotechnol* **2014**, 9, 682.

(6) Ubrig, N.; Ponomarev, E.; Zultak, J.; Domaretskiy, D.; Zólyomi, V.; Terry, D.; Howarth, J.; Gutiérrez-Lezama, I.; Zhukov, A.; Kudrynskyi, Z. R.; Kovalyuk, Z. D.; Patané, A.; Taniguchi, T.; Watanabe, K.; Gorbachev, R. V.; Fal'ko, V. I.; Morpurgo, A. F. Design of van der Waals interfaces for broad-spectrum optoelectronics. *Nat Mater* **2020**, 19, 299.

(7) Hill, J. C.; Landers, A. T.; Switzer, J. A. An electrodeposited inhomogeneous metal-insulator-semiconductor junction for efficient photoelectrochemical water oxidation. *Nat Mater* **2015**, 14, 1150.

(8) Deng, D.; Novoselov, K. S.; Fu, Q.; Zheng, N.; Tian, Z.; Bao, X. Catalysis with two-dimensional materials and their heterostructures. *Nat Nanotechnol* **2016**, 11, 218.

(9) Laskowski, F. A. L.; Oener, S. Z.; Nellist, M. R.; Gordon, A. M.; Bain, D. C.; Fehrs, J. L.; Boettcher, S. W. Nanoscale semiconductor/catalyst interfaces in photoelectrochemistry. *Nat Mater* **2020**, 19, 69.

(10) Graetzel, M.; Janssen, R. A. J.; Mitzi, D. B.; Sargent, E. H. Materials interface engineering for solution-processed photovoltaics. *Nature* **2012**, 488, 304.

(11) Zhou, H.; Chen, Q.; Li, G.; Luo, S.; Song, T.-B.; Duan, H.-S.; Hong, Z.; You, J.; Liu, Y.; Yang, Y. Interface engineering of highly efficient perovskite solar cells. *Science* **2014**, 345, 542.

(12) Vogt, P.; De Padova, P.; Quaresima, C.; Avila, J.; Frantzeskakis, E.; Asensio, M. C.; Resta, A.; Ealet, B.; Le Lay, G. Silicene: Compelling experimental evidence for graphene like two-dimensional silicon. *Phys Rev Lett* **2012**, 108, 155501.

(13) Feng, B.; Ding, Z.; Meng, S.; Yao, Y.; He, X.; Cheng, P.; Chen, L.; Wu, K. Evidence of silicene in honeycomb structures of silicon on Ag(111). *Nano Lett* **2012**, 12, 3507.

(14) Fleurence, A.; Friedlein, R.; Ozaki, T.; Kawai, H.; Wang, Y.; Yamada-Takamura, Y. Experimental evidence for epitaxial silicene on diboride thin films. *Phys Rev Lett* **2012**, 108, 245501.

(15) Li, L.; Lu, S.-Z.; Pan, J.; Qin, Z.; Wang, Y.-Q.; Wang, Y.; Cao, G.-Y.; Du, S.; Gao, H.-J. Buckled germanene formation on Pt(111). *Adv Mater* **2014**, 26, 4820.

(16) Mannix, A. J.; Zhou, X.-F.; Kiraly, B.; Wood, J. D.; Alducin, D.; Myers, B. D.; Liu, X.; Fisher, B. L.; Santiago, U.; Guest, J. R.; Yacaman, M. J.; Ponce, A.; Oganov, A. R.; Hersam, M. C.; Guisinger, N. P. Synthesis of borophenes: Anisotropic, two-dimensional boron polymorphs. *Science* **2015**, 350, 1513.

(17) Feng, B.; Zhang, J.; Zhong, Q.; Li, W.; Li, S.; Li, H.; Cheng, P.; Meng, S.; Chen, L.; Wu, K. Experimental realization of two-dimensional boron sheets. *Nat Chem* **2016**, 8, 563.

(18) Zhang, J. L.; Zhao, S.; Han, C.; Wang, Z.; Zhong, S.; Sun, S.; Guo, R.; Zhou, X.; Gu, C. D.; Yuan, K. D.; Li, Z.; Chen, W. Epitaxial growth of single layer blue phosphorus: A new phase of two-dimensional phosphorus. *Nano Lett* **2016**, 16, 4903.

(19) Zhang, J. L.; Han, C.; Hu, Z.; Wang, L.; Liu, L.; Wee, A. T. S.; Chen, W. 2D phosphorene: Epitaxial

growth and interface engineering for electronic devices. *Adv Mater* **2018**, 30, 1802207.
(20) Reis, F.; Li, G.; Dudy, L.; Bauernfeind, M.; Glass, S.; Hanke, W.; Thomale, R.; Schäfer, J.; Claessen, R. Bismuthene on a SiC substrate: A candidate for a high-temperature quantum spin Hall material. *Science* **2017**, 357, 287.
(21) Deng, J.; Xia, B.; Ma, X.; Chen, H.; Shan, H.; Zhai, X.; Li, B.; Zhao, A.; Xu, Y.; Duan, W.; Zhang, S.-C.; Wang, B.; Hou, J. G. Epitaxial growth of ultraflat stanene with topological band inversion. *Nat Mater* **2018**, 17, 1081.
(22) Shao, Y.; Liu, Z.-L.; Cheng, C.; Wu, X.; Liu, H.; Liu, C.; Wang, J.-O.; Zhu, S.-Y.; Wang, Y.-Q.; Shi, D.-X.; Ibrahim, K.; Sun, J.-T.; Wang, Y.-L.; Gao, H.-J. Epitaxial growth of flat antimonene monolayer: A new honeycomb analogue of graphene. *Nano Lett* **2018**, 18, 2133.
(23) Niu, T.; Zhou, W.; Zhou, D.; Hu, X.; Zhang, S.; Zhang, K.; Zhou, M.; Fuchs, H.; Zeng, H. Modulating epitaxial atomic structure of antimonene through interface design. *Adv Mater* **2019**, 31, 1902606.
(24) Chen, Z.; Santoso, I.; Wang, R.; Xie, L. F.; Mao, H. Y.; Huang, H.; Wang, Y. Z.; Gao, X. Y.; Chen, Z. K.; Ma, D.; Wee, A. T. S.; Chen, W. Surface transfer hole doping of epitaxial graphene using $MoO_3$ thin film. *Appl Phys Lett* **2010**, 96, 213104.
(25) Zhu, F.-F.; Chen, W.-J.; Xu, Y.; Gao, C.-L.; Guan, D.-D.; Liu, C.-H.; Qian, D.; Zhang, S.-C.; Jia, J.-F. Epitaxial growth of two-dimensional stanene. *Nat Mater* **2015**, 14, 1020.
(26) Aziza, Z. B.; Henck, H.; Pierucci, D.; Silly, M. G.; Lhuillier, E.; Patriarche, G.; Sirotti, F.; Eddrief, M.; Ouerghi, A. Van der Waals epitaxy of GaSe/graphene heterostructure: Electronic and interfacial properties. *ACS Nano* **2016**, 10, 9679.
(27) Léonard, F.; Alec Talin, A. Electrical contacts to one- and two-dimensional nanomaterials. *Nat Nanotechnol* **2011**, 6, 773.
(28) Shih, C.-J.; Wang, Q. H.; Son, Y.; Jin, Z.; Blankschtein, D.; Strano, M. S. Tuning on-off current ratio and field-effect mobility in a $MoS_2$-graphene heterostructure via Schottky barrier modulation. *ACS Nano* **2014**, 8, 5790.
(29) Allain, A.; Kang, J.; Banerjee, K.; Kis, A. Electrical contacts to two-dimensional semiconductors. *Nat Mater* **2015**, 14, 1195.
(30) Liu, Y.; Stradins, P.; Wei, S.-H. Van der Waals metal-semiconductor junction: Weak Fermi level pinning enables effective tuning of Schottky barrier. *Sci Adv* **2016**, 2, 1600069.
(31) Liu, Y.; Huang, Y.; Duan, X. Van der Waals integration before and beyond two-dimensional materials. *Nature* **2019**, 567, 323.
(32) Wang, Q.-Y.; Li, Z.; Zhang, W.-H.; Zhang, Z.-C.; Zhang, J.-S.; Li, W.; Ding, H.; Ou, Y.-B.; Deng, P.; Chang, K.; Wen, J.; Song, C.-L.; He, K.; Jia, J.-F.; Ji, S.-H.; Wang, Y.-Y.; Wang, L.-L.; Chen, X.; Ma, X.-C.; Xue, Q.-K. Interface-induced high-temperature superconductivity in single unit-cell FeSe films on $SrTiO_3$. *Chin Phys Lett* **2012**, 29, 037402.
(33) Zhang, W.-H.; Sun, Y.; Zhang, J.-S.; Li, F.-S.; Guo, M.-H.; Zhao, Y.-F.; Zhang, H.-M.; Peng, J.-P.; Xing, Y.; Wang, H.-C.; Fujita, T.; Hirata, A.; Li, Z.; Ding, H.; Tang, C.-J.; Wang, M.; Wang, Q.-Y.; He, K.; Ji, S.-H.; Chen, X.; Wang, J.-F.; Xia, Z.-C.; Li, L.; Wang, Y.-Y.; Wang, J.; Wang, L.-L.; Chen, M.-W.; Xue, Q.-K.; Ma, X.-C. Direct observation of high-temperature superconductivity in one-unit-cell FeSe films. *Chin Phys Lett* **2014**, 31, 017401.
(34) Peng, R.; Xu, H. C.; Tan, S. Y.; Cao, H. Y.; Xia, M.; Shen, X. P.; Huang, Z. C.; Wen, C. H. P.; Song, Q.; Zhang, T.; Xie, B. P.; Gong, X. G.; Feng, D. L. Tuning the band structure and superconductivity in single-layer FeSe by interface engineering. *Nat Commun* **2014**, 5, 5044.
(35) Zhang, W.; Li, Z.; Li, F.; Zhang, H.; Peng, J.; Tang, C.; Wang, Q.; He, K.; Chen, X.; Wang, L.; Ma, X.; Xue,

Q.-K. Interface charge doping effects on superconductivity of single-unit-cell FeSe films on $SrTiO_3$ substrates. *Phys Rev B* **2014**, 89, 060506

(36) Ge, J.-F.; Liu, Z.-L.; Liu, C.; Gao, C.-L.; Qian, D.; Xue, Q.-K.; Liu, Y.; Jia, J.-F. Superconductivity above 100 K in single-layer FeSe films on doped $SrTiO_3$. *Nat Mater* **2015**, 14, 285.

(37) Zhang, H.; Zhang, D.; Lu, X.; Liu, C.; Zhou, G.; Ma, X.; Wang, L.; Jiang, P.; Xue, Q.-K.; Bao, X. Origin of charge transfer and enhanced electron-phonon coupling in single unit-cell FeSe films on $SrTiO_3$. *Nat Commun* **2017**, 8, 214.

(38) Zhao, W.; Li, M.; Chang, C.-Z.; Jiang, J.; Wu, L.; Liu, C.; Moodera, J. S.; Zhu, Y.; Chan, M. H. W. Direct imaging of electron transfer and its influence on superconducting pairing at $FeSe/SrTiO_3$ interface. *Sci Adv* **2018**, 4, 2682.

(39) Shi, Z.-Q.; Li, H.; Yuan, Q.-Q.; Song, Y.-H.; Lv, Y.-Y.; Shi, W.; Jia, Z.-Y.; Gao, L.; Chen, Y.-B.; Zhu, W.; Li, S.-C. Van der Waals heteroepitaxial growth of monolayer Sb in a puckered honeycomb structure. *Adv Mater* **2019**, 31, 1806130.

(40) Kane, C. L.; Mele, E. J. Quantum spin Hall effect in graphene. *Phys Rev Lett* **2005**, 95, 226801.

(41) Liu, C.-C.; Feng, W.; Yao, Y. Quantum spin Hall effect in silicene and two-dimensional germanium. *Phys Rev Lett* **2011**, 107, 076802.

(42) Xu, Y.; Yan, B.; Zhang, H.-J.; Wang, J.; Xu, G.; Tang, P.; Duan, W.; Zhang, S.-C. Large-gap quantum spin Hall insulators in tin films. *Phys Rev Lett* **2013**, 111, 136804.

(43) Gou, J.; Kong, L.; Li, H.; Zhong, Q.; Li, W.; Cheng, P.; Chen, L.; Wu, K. Strain-induced band engineering in monolayer stanene on Sb(111). *Phys Rev Mater* **2017**, 1, 054004.

(44) Wang, G.; Pandey, R.; Karna, S. P. Atomically thin group V elemental films: Theoretical investigations of antimonene allotropes. *ACS Appl Mater Interfaces* **2015**, 7, 11490.

(45) Aktürk, O. Ü.; Özçelik, V. O.; Ciraci, S. Single-layer crystalline phases of antimony: Antimonenes. *Phys Rev B* **2015**, 91, 235446.

(46) Wu, Y.; Xu, K.; Ma, C.; Chen, Y.; Lu, Z.; Zhang, H.; Fang, Z.; Zhang, R. Ultrahigh carrier mobilities and high thermoelectric performance at room temperature optimized by strain-engineering to two-dimensional aw-antimonene. *Nano Energy* **2019**, 63, 103870.

(47) Xiao, C.; Wang, F.; Yang, S. A.; Lu, Y.; Feng, Y.; Zhang, S. Elemental ferroelectricity and antiferroelectricity in group-V monolayer. *Adv Funct Mater* **2018**, 28, 1707383.

(48) Zhao, L.-D.; Lo, S.-H.; Zhang, Y.; Sun, H.; Tan, G.; Uher, C.; Wolverton, C.; Dravid, V. P.; Kanatzidis, M. G. Ultralow thermal conductivity and high thermoelectric figure of merit in SnSe crystals. *Nature* **2014**, 508, 373.

(49) Chang, C.; Wu, M.; He, D.; Pei, Y.; Wu, C.-F.; Wu, X.; Yu, H.; Zhu, F.; Wang, K.; Chen, Y.; Huang, L.; Li, J.-F.; He, J.; Zhao, L.-D. 3D charge and 2D phonon transports leading to high out-of-plane ZT in n-type SnSe crystals. *Science* **2018**, 360, 778.

(50) Sun, Y.; Wu, S.-C.; Ali, M. N.; Felser, C.; Yan, B. Prediction of Weyl semimetal in orthorhombic $MoTe_2$. *Phys Rev B* **2015**, 92, 161107.

(51) Kresse, G.; Furthmüller, J. Efficient iterative schemes for *ab initio* total-energy calculations using a plane-wave basis set. *Phys Rev B* **1996**, 54, 11169.

(52) Blöchl, P. E. Projector augmented-wave method. *Phys Rev B* **1994**, 50, 17953.

(53) Kresse, G.; Joubert, D. From ultrasoft pseudopotentials to the projector augmented-wave method. *Phys Rev B* **1999**, 59, 1758.

(54) Perdew, J. P.; Burke, K.; Ernzerhof, M. Generalized gradient approximation made simple. *Phys Rev Lett* **1996**, 77, 3865.

(55) Krukau, A. V.; Vydrov, O. A.; Izmaylov, A. F.; Scuseria, G. E. Influence of the exchange screening parameter on the performance of screened hybrid functionals. *J Chem Phys* **2006**, 125, 224106.

(56) Grimme, S. Semiempirical GGA-type density functional constructed with a long-range dispersion correction. *J Comput Chem* **2006**, 27, 1787.